\documentstyle[epsf]{l-aa}   
\input{psfig}
\def\grs{\mbox{GRS } 1915+105}

\def\Kkmsdeg2{\mbox{ K km s}^{-1} \mbox{deg}^{2}}

\begin{document}
\thesaurus{
02      
        (02.01.2               
         02.02.1               
         02.09.1               
         02.16.1)              
08      
        (08.06.1               
         08.09.2 GRS 1915+105; 
         08.22.3)              
13      
        (13.09.6;              
         13.18.5               
         13.25.1               
         13.25.5)              
}

\title{Accretion instabilities and jet formation in GRS 1915+105}

\author{I.F. Mirabel \inst{1,2}\and V. Dhawan \inst{3}\and S. Chaty \inst{1} \and L.F. 
Rodr\'{\i}guez \inst{4} \and J. Mart\'{\i} \inst{1} \and C.R. Robinson \inst{5} \and J. 
Swank \inst{6} 
\and T.Geballe \inst{7}}
\author{I.F. Mirabel \inst{1,2}\and V. Dhawan \inst{3}\and S. Chaty \inst{1}\and L.F. 
Rodr\'{\i}guez \inst{4}\and J.Mart\'{\i} \inst{1}\and C.R. Robinson\inst{5}\and J. Swank 
\inst{6}\and T.R. Geballe \inst{7}}

\institute{
Service d'Astrophysique, CEA/DSM/DAPNIA/SAp, Centre d'\'etudes de Saclay, F-91191 
Gif-sur-Yvette Cedex, France
\and
Instituto de Astronom\'\i a y F\'\i sica del Espacio. C.C.67, Suc. 28, 1428 Buenos Aires, 
Argentina
\and
National Radio Astronomy Observatory, Socorro, NM 87801, USA
\and
Instituto de Astronom\'{\i}a, UNAM, Morelia, Michoac\'an 58090, Mexico
\and
Marshall Space Flight Center, Space Science Laboratory, ES84, Huntsville, AL 35812, USA
\and 
Goddard Space Flight Ctr., Code 666, Greenbelt, MD 20771, USA
\and 
Joint Astronomy Centre, Hawaii Headquarters, 660 N. A'ohoku Place, Hilo, HI 96720, USA}

\date{Received October 24, 1997; accepted November 10, 1997}

\maketitle

\markboth{I.F. Mirabel et al.: Accretion instabilities and jet formation in GRS 
1915+105}{}

\begin{abstract}
We report simultaneous observations in the X-ray, infrared, and radio wavelengths of the 
galactic superluminal source GRS 1915+105. During episodes of rapid disappearance and 
follow up replenishment of the inner accretion disk evidenced by the X-ray oscillating 
flux, we observe the ejection of relativistic plasma clouds in the form of synchrotron 
flares at infrared and radio wavelengths. The expelled clouds contain very energetic 
particles with Lorentz factors of $\sim$10$^3$, or more. These ejections can be viewed as 
small-scale analogs of the more massive ejecta with relativistic bulk motions that have 
been previously observed in GRS 1915+105. 


\end{abstract}

\section{Introduction}

The discovery of superluminal jets (Mirabel \& Rodr\'\i guez 1994) in the black hole 
X-ray transient source GRS 1915+105 has opened
the possibility of studying phenomena in our Galaxy that until
recently were believed to be restricted to
distant quasars and a few active galactic nuclei. 
In particular, it has been realized that since
the characteristic dynamical times in the flow of matter onto a black hole  
are proportional to its mass, the events with 
intervals of minutes in a microquasar could correspond
to analogous phenomena with duration of thousands of years in a quasar of 
10$^9$ M$_{\odot}$, which is much longer than a human life-time.
Therefore, the variations with minutes of duration observed in GRS 1915+105 
in the radio, IR, and X rays
sample phenomena that we have not been able to observe in quasars.

X-ray observations of GRS 1915+105 with the Rossi X-Ray Timing Explorer (RXTE) revealed 
remarkable quasi-periodic oscillations (QPOs) that are believed to arise in the accretion 
disk around a black hole of stellar mass (Greiner et al. 1996, Morgan et al. 1997; 
Belloni et al. 1997a,b; Chen et al. 1997). Among the diversity of QPOs observed in GRS 
1915+105 there is a class of recurrent episodes with amplitude variations of 
$\sim$10$^{39}$ erg s$^{-1}$ (at a distance of 12.5 kpc) in the X-rays, in time scales of 
one minute to tens of minutes. These QPOs have been attributed to the rapid disappearance 
and slower replenishment of the inner region of the accretion disk (Belloni et al. 
1997a,b). If the accretion is an advection-dominated-flow 
(Narayan et al. 1997), the disappearing mass will go quietly though the horizon of the 
black hole. Furthermore, if a fraction of the mass of the accretion disk were blown away, 
one should see synchrotron emission from expanding clouds in the radio, and perhaps 
shorter wavelengths.

Although previous observations have shown flares with similar periodicities in the radio 
(Pooley \& Fender, 1997; Rodr\'\i guez \& Mirabel, 1997a) and infrared (Fender et al. 
1997), no truly simultaneous observations in the X-ray, infrared, and radio wavelengths 
have been reported so far. The observations reported here firmly establish the genesis of 
expanding clouds of relativistic plasma when GRS 1915+105 recovers from large amplitude 
dips in the X-ray flux.

\section{Multiwavelength observations and light curves}

In this Letter we report observations made on May 15 and Sept 9, 1997. On May 15 at the 
UT intervals 6-10 h and 12-16 h, we used the 27 antennae of the Very Large Array (VLA) 
split in three sub-arrays to allow simultaneous observations at 2, 3.6, and 6 cm. On Sep 
9, 17 antennae of the VLA were available for this project, and the observations were done 
at 3.6 cm only.   

The infrared observations were carried out with the United Kingdom Infrared Telescope 
(UKIRT). On each date, sixty images with the IRCAM3 256$\times$256 InSb detector array, 
with a K-band filter of 0.37$\mu$m width centered at 2.2$\mu$m, were obtained 
continuously every 1 min for an interval of $\sim$1.1 h. The 2.2$\mu$m flux of $\grs$ on 
time bins of 1 min was computed using relative photometry with several stars in the field 
and calibrated with well known standard stars. The infrared fluxes were dereddened 
assuming A$_K$=3.3 mag (Chaty et al. 1996).

The X-ray observations were done with the PCA on RXTE. It consists of five nearly 
identical proportional counter units with detectors sensitive to X-rays between 2 and 60 
keV.\\ 

The results for May 15 are shown in Figure 1. The 
RXTE observations exhibit large amplitude variations with sudden drops of the flux into 
dips of $\sim$5000 counts s$^{-1}$ that last for at least 40 min. The recovery of the 
X-ray flux from the dips sometimes ends in sharp peaks that reach $\sim$10$^{-7}$ erg 
cm$^{-2}$ s$^{-1}$ (2 mJy) with typical durations of $\sim$25 s. These rebounds of the 
X-ray flux are followed by smaller amplitude oscillations of $\sim$60 s period that last 
tens of minutes. At the time of an X-ray gap around 14 h UT, an infrared flare of 12 mJy 
was observed. The rise of the X-ray flux observed after 14.25 h UT suggests a dip during 
that gap. Interpolating the pattern, a sharp peak and sequences of oscilations would have 
occured at $\sim$13.7 h UT. 

\begin{figure*}
\plotfiddle{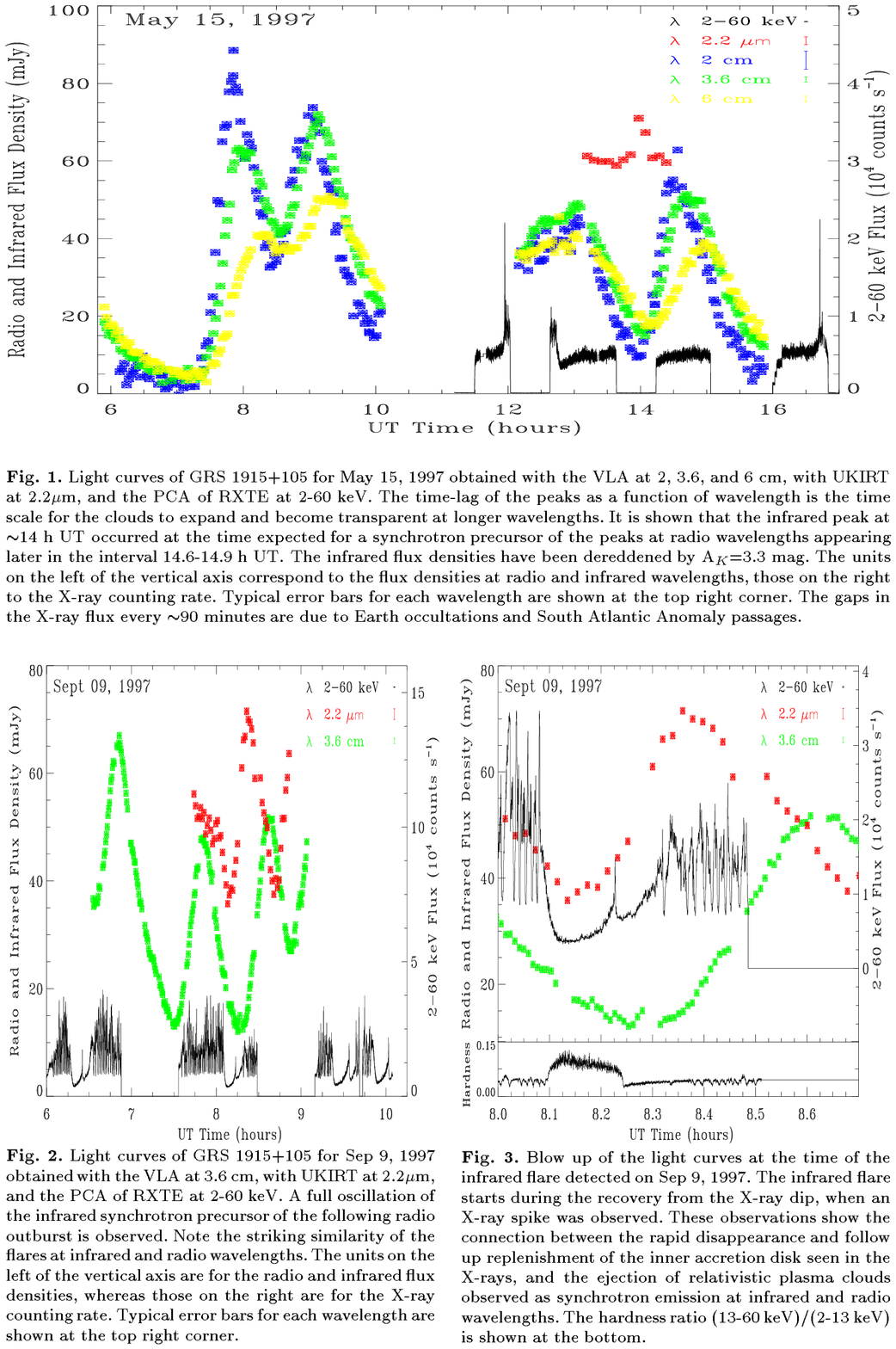}{21.0cm}{0}{100}{100}{-260}{-120}  \label{GRScoord}
\end{figure*}

The quasi-periodic oscillations in the radio flux are typical of the optically thick 
radio core state, usually observed when the source oscillates between 10 and 100 mJy. A 
time-shift of the flux peaks, with the short wavelengths peaking first, is clearly 
observed. 

The results for Sep 9 are shown in Figure 2. The PCA 
count rates show large amplitude oscillations ($\sim$10$^{39}$ erg s$^{-1}$) every 
$\sim$50 s for intervals of tens of minutes, which are followed by sudden drops into
dips. The recovery from the dips is slower than the time of decay and lasts a few 
tens of minutes. In the X-ray light curves of May 15 and Sep 9 we note the following 
similarities: 1) both the $\sim$60 s oscillations on May 15 and the $\sim$50 s 
oscillations of Sep 9 disappear during the dips, and 2) the low intensity value in both 
cases is about 5000 counts s$^{-1}$, which is well above the background rate of 100 
counts s$^{-1}$. However, there are obvious differences between the light curves in the 
two dates. 1) On May 15 the X-ray dips lasted longer than on Sep 9. 2) On May 15 the 
infrared flare had smaller amplitude and duration than the radio flare, whereas on Sep 9 
they had similar amplitude and duration. 3) On May 15 the time-shift between the infrared 
peak and the following peak at 3.6 cm is 40 min, 
on Sep 9 it is 16 min. 

The relation between the X-ray, infrared, and radio light curves for Sep 9 is shown in 
more detail in Figure 3. The X-ray spectral index 
(13-60 kev)/(2-13 keV) suddenly increased at the beginning of the dip. 
Despite the 10$^4$ order of magnitude difference in wavelength, the infrared flare and 
the follow up radio flare are strikingly similar in amplitude, duration, and slope of the 
increasing and decreasing fluxes. The infrared flare seems to start during the recovery 
from the X-ray dip at $\sim$8.23 h UT, when the X-ray spectrum softened again and an 
X-ray spike was observed. The rise of the infrared flux to maximum value continues for a 
few minutes after the 50 s oscillations in the X-ray flux appeared again. 

\section{Infrared synchrotron precursors}

The radio emission from GRS~1915+105 during 1997
May 15 shows a behavior consistent with that expected for
synchrotron emission from an adiabatically expanding cloud
(van der Laan 1966).
In all four bursts observed in the period of 6 to 16 h UT, the light curves
at 6, 3.6, and 2-cm
reach their peaks with the characteristic
wavelength-dependent delay (the short wavelengths peaking first).
These delays rule out the possibility that the observed events were
produced by time-variable free-free opacity, as proposed by Rodr\'\i guez \&
Mirabel (1997a) to explain sinusoidal variations observed in 1995, since in this
model no wavelength-dependent delays are expected. When the formation of clouds is faster 
than the decay of the flux, consecutive radio events blend together and the integrated 
emission from GRS 1915+105 may appear with a sinusoidal modulation (Rodr\'\i guez \& 
Mirabel, 1997a). 

In the van der Laan (1966) model, 
we can estimate $p$, the energy spectral index of the relativistic
electrons ($N(E) \propto E^{-p}$), from the equation that relates the observed maximum 
flux density, $S_{m,\lambda}$, at a given wavelength, i.e,   
$S_{m,\lambda} \propto \lambda^{-(7p+3)/(4p+6)}$. For this estimate we use the last burst 
observed on May 15 between 14 and 16 h UT, since it appears to be more isolated than the 
others.
Since $S_{m,6cm}$ = 39 mJy, and $S_{m,3.6cm}$ = 51 mJy,
we obtain $p~\simeq$ 0.

The ejection
is defined to occur at $t = t_0$. The time $t_{m,\lambda}$ since ejection 
in which the light curve at a given wavelength reaches maximum
is given by $t_{m,\lambda} \propto \lambda^{(p+4)/(4p+6)}$. With $p~\simeq$ 0 we obtain 
$t_{6cm }/ t_{3.6cm}$ = 1.4. Furthermore, from the data in Figure 1, we
have $t_{6cm } + t_0$ = 14.95 h UT and $t_{3.6cm } + t_0$ = 14.70 h UT.
Therefore,
$t_{6cm } - t_{3.6cm}$ = 0.25 h, and 
$t_{6cm } \simeq$ 0.9 h. We then find that the ejection of the plasma occurred at $t_0$ = 
14.05 h UT.

In this simple model, the UT time of maximum flux density at a given wavelength,
for the particular event analyzed above, is given 
by $t_{m,\lambda}(UT~h) = 14.05 + 0.9 (\lambda / 6cm)^{(2/3)}.$ Then, the maximum flux 
density at short wavelengths (i.e. the near
infrared) are observed very shortly after the ejection.
It is only in the radio wavelengths that significant time delays occur.
We note that the 2.2$\mu$m emission indeed
reached maximum near 14.0 h UT.
This result strongly supports the interpretation of the
IR burst on May 15 as the synchrotron precursor of the radio bursts.
To our knowledge, this is the first time in X-ray binaries
that a clear association is established between an IR synchrotron precursor
and the follow-up radio bursts.

\section{Jet formation during the X-ray dips}

Figure 3 shows that on Sep 9 the infrared flare started during the rise of the X-ray flux 
from the dip, when a softenning of the spectrum and an X-ray peak were observed. Since at 
infrared wavelengths the ejected plasma becomes transparent very shortly after the 
ejection, the rise of the infrared flux to maximum indicates that the injection of 
relativistic particles lasted $\sim$10 min, or less.

If we assume that the synchrotron emitting plasma expands at $\sim$0.2$c$,
as observed in a larger scale ejection (Mirabel \& Rodr\'\i guez 1994), 15 minutes after 
the ejection the clouds have dimensions of 
$\sim10^{13}$ cm. Using this dimension and the infrared and radio flux density values
measured on Sep 9, with equipartition arguments we estimate a brightness temperature of 
10$^{12}$ K, a magnetic field of 16 G, an equipartition energy in relativistic electrons 
of $5 \times 10^{39}$ erg, and a typical synchrotron luminosity integrated from 2.2$\mu$m 
to 6 cm of the order of 10$^{36}$ erg s$^{-1}$. These parameters are comparable to those 
derived by Fender et al. (1997) from observations on other epochs. 
If the electrons that produce the synchrotron radiation have a representative Lorentz 
factor of $\sim$10$^3$, and there is one proton per electron, the minimum mass of the 
clouds that are ejected every few tens of minutes is $\sim$ 10$^{19}$ g. 

For the parameters of these clouds the energy losses are dominated by adiabatic expansion 
and the clouds should have a lifetime of about 1 h, as observed here. An infrared jet as 
the one reported by Sams, Eckart \& Sunyaev (1996), which reached distances of $\sim$0.2 
arcsec (2500 AU at 12.5 kpc) requires very energetic particles, and a re-acceleration 
mechanism. We point out that radio blobs with bulk motions of 0.92$c$ and lifetimes of 
several weeks have been observed (Rodr\'\i guez \& Mirabel, 1997b).

\section{Conclusions}

1) At times of quasi-periodic oscillations of large amplitude in the X-rays, relativistic 
plasma clouds emanate from the X-ray binary. The time shift of the emission at radio 
wavelengths is consistent with synchrotron radiation from gas in adiabatic expansion. 

\noindent 2) The association of infrared synchrotron precursors to follow-up bursts at 
radio wavelengths is firmly established. 

\noindent 3) By the simultaneous observations reported here, we definitively confirm that 
the infrared and radio flares are associated to X-ray dips, 
as suggested by Fender et al (1997) and Pooley \& Fender (1997). The onset of the 
ejection takes place during the recovery from the X-ray dip of hard (13-60 keV)/(2-13 
keV) spectrum, at the time of a spike in the X-rays. 

\noindent 4) The expanding clouds of plasma reported here can be viewed as small-scale 
analogs of the more massive ejections associated to large radio outbursts (Rodr\'\i guez 
\& Mirabel, 1997b). The equipartition energy in relativistic electrons in the mini-ejecta 
is of order of $5 \times 10^{39}$ erg, namely, $\sim$10$^3$ times less energetic than in 
major ejecta. However, the genesis of these smaller scale clouds is  much more frequent, 
and may represent an important phenomenon in the long term evolution of the binary and 
its environment. 

\noindent 5) The multiwavelength variations with minutes of duration in GRS 1915+105 show 
how the sudden disappearance of the inner accretion disk around a black hole of stellar 
mass triggers the formation of relativistic expanding clouds. Analogous phenomena in 
quasars with black holes of 10$^9$ M$_{\odot}$ would require thousands of years of 
observations. Therefore, microquasars in our own Galaxy can provide new information on 
the physics of accretion disks around black holes and are thus useful to gain insight 
into the relativistic ejections seen elsewhere in the Universe.

\begin{acknowledgements} We thank M. Robberto and S. Leggett for the UKIRT service 
observations on May 15, and R.A. Garc\'\i a for assistance with the graphic displays. We 
also thank S. Eikenberry and R. Blandford for communicating results prior to publication. 
The NRAO is a facility of the N.S.F. under cooperative agreement with Associated 
Universities Inc. UKIRT is operated by the Joint Astronomy Centre on behalf of the U.K. 
Particle Physics and Astronomy Research Council. J.M. is supported by the Spanish 
MEC.\end{acknowledgements}


\end{document}